\documentclass[a4paper]{revtex4}
\usepackage{epsfig}

\newcommand{\be}{\begin{equation}}
\newcommand{\ee}{\end{equation}}
\newcommand{\bea}{\begin{eqnarray}}
\newcommand{\eea}{\end{eqnarray}}
\newcommand{\bean}{\begin{eqnarray*}}
\newcommand{\eean}{\end{eqnarray*}}

\newcommand{\gapproxeq}{\lower
.7ex\hbox{$\;\stackrel{\textstyle >}{\sim}\;$}}
\newcommand{\lapproxeq}{\lower
.7ex\hbox{$\;\stackrel{\textstyle <}{\sim}\;$}}

\begin{document}

\bibliographystyle{unsrt}

\title{\bf A study of the Okubo-Zweig-Iizuka rule violations in $\eta_c\to VV$ }

\author{Qiang Zhao$^{1,2}$ }

\affiliation{1) Institute of High Energy Physics, Chinese Academy
of Sciences, Beijing, 100049, P.R. China}

\affiliation{2)Department of Physics, University of Surrey,
Guildford, GU2 7XH, United Kingdom}

\date{\today}

\begin{abstract}

We discuss the Okubo-Zweig-Iizuka (OZI) rule violation effects in
$\eta_c\to VV$ in light of the new data from BES Collaboration. In
particular, a possible non-vanishing branching ratio for
$\eta_c\to \omega\phi$ provides a hint of the degrees of OZI
violations based on a recent factorization proposed for charmonium
hadronic decays. The violation mechanism is studied via an
intermediate meson exchange model. The results are consistent with
the experimental observations.

\end{abstract}

\maketitle


The new data from BES Collaboration for $\eta_c\to
VV$~\cite{bes-eta-c}, where $V$ denotes vector meson, revive the
question about the role played by non-perturbative QCD processes.
In particular, an upper limit for $\eta_c\to \omega\phi$ branching
ratio, which is the same order of magnitude as that for $\eta_c\to
\phi\phi$, initiates renewed interests in possible
OZI-rule~\cite{ozi} violations in charmonium hadronic decays.
Since the flavor wavefunctions of $\omega$ and $\phi$ are nearly
ideally mixing, i.e. $\phi$ meson is a pure $s\bar{s}$ and
$\omega$, pure $n\bar{n}\equiv (u\bar{u}+d\bar{d})/\sqrt{2}$, in
the limit where the OZI-rule applies, the branching ratio for the
doubly OZI disconnected (DOZI) process $\eta_c\to \omega\phi$ is
expected to vanish, or at least be much smaller than the singly
OZI disconnected (SOZI) processes such as $\eta_c\to
K^*\bar{K^*}$, etc.  Therefore, a possible non-negligible
branching ratio for $\eta_c\to \omega\phi$ would be a direct
evidence for the break-down of pQCD in the low-lying charmonium
hadronic decays, for which the dynamical reason needs to be
understood. Some theoretical attempts towards a dynamical
understanding of $\eta_c\to VV$ can be found in the
literature~\cite{chernyak,anselmino-90,anselmino-94,benayoun,feldmann,zhou-ping-zou}.

A dynamical prescription for the OZI violations in the charmonium
hadronic decays can be the intermediate meson exchange
mechanism~\cite{lu-zou-locher,li-bugg-zou}. As proposed in
Ref.~\cite{zzm}, the DOZI processes could be dual with the
intermediate meson exchanges, which is in analogue with Geiger and
Isgur's study~\cite{isgur-geiger} of OZI violation mechanisms. For
$J/\psi\to V f_0$~\cite{zzm}, we found that the OZI violations via
the intermediate meson exchanges accounted for the branching ratio
patterns arising from $J/\psi\to V f_0^i$, where $i=1,2,3$ denote
$f_0(1710)$, $f_0(1500)$ and $f_0(1370)$. Similarly, this
prescription can be applied to $\eta_c\to VV$, where the
intermediate meson exchanges could be a natural explanation for a
non-vanishing $\eta_c\to \omega\phi$ decay.

In this letter, we provide a coherent analysis for $\eta_c\to VV$
based on a recently proposed factorization
scheme~\cite{zhao-chi-c}. By determining the OZI-violation
parameter defined in the factorization, we will be able to predict
the range of branching ratio magnitude for $\eta_c\to \omega\phi$.
We then investigate this DOZI process by applying the intermediate
meson exchange mechanism. Conclusion and some discussions will be
given in the end.

In the factorization scheme~\cite{zhao-chi-c}, the DOZI process is
distinguished from the SOZI ones by the gluon counting rule.
Namely, we assume that the transition amplitude of DOZI process
has a strength of $r$ times that of SOZI due to additional gluon
exchange. We also introduce the SU(3) flavor breaking parameter
$R$, of which its deviation from unity reflects the change of
couplings due to the mass difference between $u/d$ and $s$. The
third parameter $t$ is introduced to describe the coupling
difference between gluon-$q\bar{q}$ and gluon-glueball. For
$\eta_c\to VV$, the transitions will not involve parameters $t$
since we treat $\eta_c$ as a pure $c\bar{c}$ state. Finally, the
gluon-$q\bar{q}$ coupling is defined by parameter $g_0$. As shown
in Ref.~\cite{zhao-chi-c}, $g_0$ is a stable quantity insensitive
to $\chi_{0,2}\to VV$ and $PP$. This property reflects that the
gluon-$q\bar{q}$ coupling has a universal value at the same energy
region.

Based on the above simple scheme, the transition amplitudes for
$\eta_c\to VV$ can be expressed as
\bea
\label{isospin-1} \langle \phi\phi | \hat{V}_{gg}| \eta_c\rangle
&=& g_0^2 R^2 (1+r)   \nonumber\\
\langle \omega\omega | \hat{V}_{gg}| \eta_c\rangle
&=& g_0^2(1+2r)  \nonumber\\
\langle \omega\phi | \hat{V}_{gg}| \eta_c\rangle &=& g_0^2 r R
\sqrt{2}
\nonumber\\
\langle K^{*+}K^{*-}|  \hat{V}_{gg}| \eta_c\rangle
&=& g_0^2 R \nonumber\\
\langle \rho^+\rho^- | \hat{V}_{gg}| \eta_c\rangle &=& g_0^2  \ .
\eea
where $\hat{V}_{gg}$ is the $\eta_c\to gg\to (q\bar{q})(q\bar{q})$
potential. The amplitudes for other charge combinations of
$K^*\bar{K^*}$ and $\rho\rho$ are the same.

A commonly used form factor is adopted in the calculation of the
partial decay widths:
\be
{\cal F}^2({\bf p})=p^{2l}\exp(-{\bf p}^2/8\beta^2) \ ,
\ee
where ${\bf p}$ and $l$ are the three momentum and relative
angular momentum of the final-state mesons, respectively, in the
$\eta_c$ rest frame. We adopt $\beta=0.5$ GeV, which is the same
as in
Refs.~\cite{close-amsler,close-kirk,close-zhao-f0,zhao-chi-c}.
Such a form factor will largely account for the size effects from
the spatial wavefunctions of the initial and final state mesons.

In Ref.~\cite{bes-eta-c}, branching ratios for $\eta_c\to
\rho\rho$, $K^*\bar{K^*}$, $\phi\phi$ are measured with improved
error bars compared with PDG~\cite{pdg2004}, and upper limits for
$\eta_c\to \omega\omega$ and $\omega\phi$ are estimated. In
particular, the upper limit, $BR_{\eta_c\to \omega\phi}< 1.3\times
10^{-3}$, is the same order of magnitude as $BR_{\eta_c\to
\phi\phi}=(2.5\pm 0.5\pm 0.9)\times 10^{-3}$, which implies that
large OZI violations might exist in $\eta_c\to VV$.

Adopting the data of Ref.~\cite{bes-eta-c}, we make two fittings
in the factorization scheme. The parameters are listed in
Table~\ref{tab-1} and the fitting results are presented in
Table~\ref{tab-2}. In Fit-I, the three parameters, $r$, $R$ and
$g_0$, are determined by fitting the branching ratios for
$\eta_c\to \rho\rho$, $K^*\bar{K^*}$ and $\phi\phi$. Since there
are three parameters to be fitted by three experimental data, the
negligibly-small $\chi^2$ reflects the fact that one can also
determine the parameters by solving the three variable equations.
Interestingly, large errors are found with the OZI violation
parameter $r$. With these fitted parameters the predicted
branching ratios for $\eta_c\to \omega\omega$ are nearly as large
as two times the experimental upper limit. This reflects the lack
of information about the parameter correlations among the three
channels: $\rho\rho$, $K^*\bar{K^*}$ and $\phi\phi$. For instance,
the decay of $\eta_c\to \rho\rho$ only involves $g_0$, thus, can
be directly determined by the branching ratio without
interferences with the $K^*\bar{K^*}$ and $\phi\phi$ channels. Due
to this, the price to pay for a ``precise" fit for $\rho\rho$,
$K^*\bar{K^*}$ and $\phi\phi$ is that the predictions for
$\omega\omega$ and $\omega\phi$ channels will bear much larger
uncertainties though they can still be regarded as in agreement
with the data (see Fit-I of Table~\ref{tab-2}).

In order to put more constraints on the determination of the
parameters and examine their sensitivity to the experimental upper
limits for $\eta_c\to\omega\omega$ and $\omega\phi$, we set their
branching ratios to be $BR_{\eta_c\to \omega\omega}=(3.1\pm
3.1)\times 10^{-3}$ and $BR_{\eta_c\to \omega\phi}=(0.6\pm
0.6)\times 10^{-3}$, i.e. assuming that the most probable values
are at the half values of their upper limits.  We then obtain
Fit-II in Table~\ref{tab-1} and \ref{tab-2} with a reasonably
small $\chi^2$, i.e. $\chi^2=0.5$. The fitted branching ratios for
$\eta_c\to \phi\phi$ and $\rho\rho$ are found smaller than the
experimental central values, while that for $K^*\bar{K^*}$ turns
out to be larger though all these are well within the uncertainty
of the data. Nevertheless, we find that the fitted values for
$\omega\omega$ and $\omega\phi$ are now well below the
experimental upper limits.

Compared with Fit-I, the uncertainties for $r$ in Fit-II are
improved; $g_0$ does not experience significant changes; and the
SU(3) flavor symmetry seems to be better respected. This feature
not only confirms that the new data from BES are more consistent
with the expectations based on the SU(3) flavor
symmetry~\cite{bes-eta-c}, but also suggests that the OZI
violations are phenomena different from the SU(3) flavor symmetry
breaking. Moreover, the value, $g_0\sim 0.36$ GeV$^{1/2}$, is
stable and close to that found in $\chi_{c0,2}\to VV$ ($g_0\simeq
0.25$ GeV$^{1/2}$)~\cite{zhao-chi-c}. Since $g_0^2$ is
proportional to the value of the $c\bar{c}$ wavefunction at its
origin, these results indicate some dynamical similarities between
$\eta_c\to VV$ and $\chi_{c0,2}\to VV$, and can be regarded as a
consistent test of the factorization scheme.

To show the uncertainties bearing with the predictions for
$\eta_c\to \omega\phi$ and $\omega\omega$, we also present the
root mean square errors for these two channels in both fittings.

\begin{figure}
\begin{center}
\epsfig{file=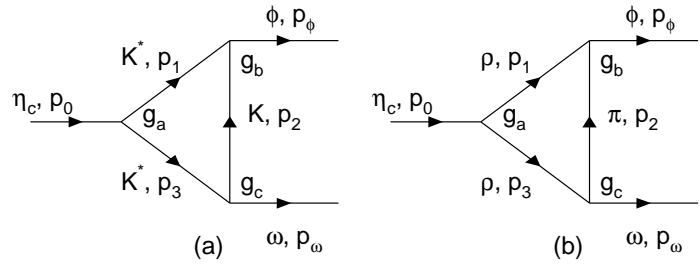, width=12cm,height=6.cm} \caption{Schematic
pictures for the decays of $\eta_c\to\omega\phi$ via (a)
$K^*\bar{K^*}$ and (b) $\rho\rho$. } \protect\label{fig-1}
\end{center}
\end{figure}

Note that $\eta_c$ has large branching ratios to $\rho\rho$ and
$K^*\bar{K^*}$ in its two-body decays. These transitions can occur
via SOZI processes, while $\eta_c\to \omega\phi$ can only occur
via DOZI ones. This is a natural explanation for the branching
ratio suppression for $\eta_c\to \omega\phi$ in comparison with
that for $\eta_c\to K^*\bar{K^*}$ and $\rho\rho$. However, note
that both $\omega$ and $\phi$ have sizeable couplings to
$K^*\bar{K}+c.c.$ and $\rho\pi$~\cite{pdg2004}. It implies that
the $K^*\bar{K^*}$ (and/or $\rho\rho$) rescattering into
$\omega\phi$ via kaon (and/or pion) as shown by Fig.~\ref{fig-1},
could lead to sizeable contributions from the DOZI processes and
be the major source for the OZI violations.

To estimate the meson exchange contributions to $\eta_c\to
\omega\phi$, we evaluate the meson loop transitions and express
the transition amplitude as follows:
\be
M_{fi} = \int \frac{d^4 p_2}{(2\pi)^4}\delta^4(P_0-P_\phi
-P_\omega) \sum_{K^* pol.}\frac{T_a T_b T_c}{a_1 a_2 a_3} {\cal
F}(p_2^2) \ ,
\ee
where the vertex functions are
\bea
T_a &\equiv & \frac{i g_a}{M_0}\epsilon_{\alpha\beta\gamma\delta}
p_1^\alpha \epsilon_1^\beta p_3^\gamma \epsilon_3^\delta ,
\nonumber\\
T_b &\equiv & \frac{i g_b}{M_\phi}\epsilon_{\mu\nu\xi\tau}
p_\phi^\mu \epsilon_\phi^\nu p_1^\xi \epsilon_1^\tau , \nonumber\\
T_c &\equiv & \frac{i
g_c}{M_\omega}\epsilon_{\lambda\iota\kappa\sigma}p_\omega^\lambda
\epsilon_\omega^\iota p_3^\kappa \epsilon_3^\sigma ,
\eea
where $g_a$, $g_b$ and $g_c$ are coupling constants at the meson
interaction vertices (see Fig.~\ref{fig-1}); Note that the tensor
part of the vector meson propagator will not contribute. The
four-vectors, $P_0$, $P_\phi$ and $P_\omega$, are momenta for the
initial $\eta_c$ and final state $\phi$ and $\omega$ mesons, while
$p_1$, $p_2$ and $p_3$ are four momenta for the intermediate
mesons, respectively. Quantities, $a_1=p_1^2-m_1^2$,
$a_2=p_2^2-m_2^2$ and $a_3=p_3^2-m_3^2$, are the denominators of
the propagators of intermediate mesons.

By applying the Cutkosky rule to the loop integration, we can
reduce the transition amplitude to be
\be
M_{fi} = \frac{-i g_a g_b g_c |{\bf p}_3|}{128 \pi^2 M_\phi
M_\omega M_0^2 } \epsilon^{\lambda\iota\kappa\delta}
P_{\omega\lambda}\epsilon_{\omega\iota} P_{\phi\kappa}
\epsilon_{\phi\delta} {\cal I} ,
\ee
with
\be
\label{inte} {\cal I}\equiv\int \frac{p_2^2
[2p_2^2+2(P_\phi^2-p_1^2) - P_\phi\cdot P_\omega ] {\cal
F}(p_2^2)}{p_2^2-m_2^2} d\Omega \ .
\ee
In the above the integration is over the azimuthal angles of the
momentum ${\bf p}_1$ respect to the momentum of the final state
$\phi$ meson, and $p_2^2=(P_\phi-p_1)^2=M_\phi^2+m_1^2-2 E_\phi
E_1 +2 |{\bf P}_\phi||{\bf p}_1|\cos \theta$. The coupling
constant $g_a$ can be determined by the experimental data for
$\eta_c\to K^*\bar{K^*}$, which is an SOZI process and one of the
largest channels in the $\eta_c$ decay:
\be
g_a^2=\frac{4\pi M_0^2}{|{\bf p}_1|^3}\Gamma_{\eta_c\to
K^*\bar{K^*}} \ ,
\ee
where $\Gamma_{\eta_c\to K^*\bar{K^*}}=10.4\times
10^{-3}\Gamma_{tot}=0.18$ MeV.

Since the exchanged particles are correlated at those three
vertices and the charge combination factor has been included in
the derived coupling constant $g_a$, we determine the $g_b $ and
$g_c$ by the SU(3) relation in comparison with
$g_{\omega\rho^0\pi^0}$:
\bea
g_b & = & g_{\phi K^{*0} \bar{K^0}} =  g_{\phi K^{*+}K^-}
=\sqrt{2}g_{\omega\rho^0\pi^0}=\sqrt{2} g_{\omega\rho^+\pi^-},
\nonumber \\
g_c &= & g_{\omega K^{*0} \bar{K^0}} =  g_{\omega \bar{K^{*0}}
K^0} = g_{\omega K^{*+} K^-}= g_{\omega K^{*-} K^+} =
g_{\omega\rho^0\pi^0},
\eea
with $g^2_{\omega\rho^0\pi^0}\simeq 84$ determined in vector meson
dominance model in $\omega\to \pi^0 e^+ e^-$~\cite{pdg2004}.
However, the SU(3) flavor symmetry is generally broken. For
example, the SU(3) flavor symmetry gives $g_{\phi\rho\pi}=0$. But
we know that the decay of $\phi\to \rho\pi$ has sizeable branching
ratios~\cite{pdg2004}. To include the contributions from
Fig.~\ref{fig-1}(b), we adopt $BR_{\phi\to\rho\pi+
\pi^+\pi^-\pi^0}=15.4\%$ as an upper limit to derive
$g_{\phi\rho\pi}$.

We also note that the application of the Cutkosky rule to the
meson loop integration for $\eta_c\to \omega\phi$ requires a
sufficient consideration of the non-locality of the $VVP$
couplings in the final state. As shown by Eq.~(\ref{inte}), the
integrand has a power of $p_2^2$, which suggests that the meson
loop integral may not be necessarily suppressed compared with the
tree diagram without form factor. Therefore, a form factor to take
care of the non-locality of the two $VVP$ vertices are needed.
This issue is correlated to the off-shell effects from the
exchanged light pseudoscalar meson. Following the argument of
Ref.~\cite{friman-soyeur}, we adopt a dipole form factor to
account for the non-locality for the two $VVP$ vertices and
off-shell effects from the exchanged meson. This should be a
natural treatment for the intermediate meson exchange model. As
follows, we will list the integration results for three cases: i)
with no form factor; ii) with a monopole form factor; and iii)
with a dipole form factor. We then concentrate on the numerical
results from (iii) due to the reason mentioned above.

i) With no form factor, i.e. ${\cal F}(p_2^2)=1$, the integral
becomes:
\be
{\cal I}= \frac{8\pi |{\bf P}_\phi||{\bf p}_1|
}{C}\left[2-\frac{2(A-B)C}{B^2}
-\frac{(A-B)(C-B)}{B^3}\ln\frac{1-B}{1+B}\right],
\ee
where the factors are
\bea
A & \equiv  & \frac{2|{\bf P}_\phi||{\bf
p}_1|}{M_\phi^2+m_1^2-2E_\phi E_1} \nonumber\\
B & \equiv & \frac{2|{\bf P}_\phi||{\bf
p}_1|}{M_\phi^2+m_1^2-2E_\phi E_1-m_2^2} \nonumber\\
C &\equiv & \frac{4|{\bf P}_\phi||{\bf p}_1|}{4M_\phi^2-4E_\phi
E_1-P_\phi\cdot P_\omega} .
\eea

ii) With a monopole form factor, i.e. ${\cal
F}(p_2^2)=(\Lambda^2-m_2^2)/(\Lambda^2-p_2^2)$, where $\Lambda$ is
the cut-off energy, the integral becomes:
\bea
{\cal I} &= & 4\pi
(\Lambda^2-m_2^2)\left[2+\frac{D(A-B)(C-B)}{ABC(B-D)}\ln\frac{1-B}{1+B}+
\frac{B(A-D)(C-D)}{ACD(B-D)}\ln\frac{1-D}{1+D} \right],
\eea
where $A$, $B$, and $C$ are the same as defined in (i), and
\be
D \equiv  \frac{2|{\bf P}_\phi||{\bf p}_1|}{M_\phi^2+m_1^2-2E_\phi
E_1-\Lambda^2} .
\ee

iii) With a dipole form factor, i.e. ${\cal
F}(p_2^2)=[(\Lambda^2-m_2^2)/(\Lambda^2-p_2^2)]^2$, the integral
becomes
\bea
{\cal I} & = & 4\pi \frac{BD(\Lambda^2-m_2^2)^2}{ACD_s} \left[
\frac{2(C-D)(A-D)}{D(B-D)(1-D^2)} + \frac{(A-B)(C-B)}{B(B-D)^2}
\ln{\frac{1-B}{1+B}}\right.\nonumber\\
&&\left.+\left(\frac{AC}{BD^2}+\frac{(A-B)(B-C)}{B(B-D)^2}\right)\ln
\frac{1-D}{1+D}\right] \ ,
\eea
where $A$, $B$, $C$ and $D$ are the same as defined in (i) and
(ii), and $ D_s\equiv M_\phi^2+m_1^2-2E_\phi E_1-\Lambda^2$.
Meanwhile, to test the sensitivity of the results to the cut-off
energies, we adopt $\Lambda=0.7, \  1.0$  and $1.2$ GeV as the
lower and upper limit in the calculations.

With the above parameters, the intermediate meson exchanges of
$K^*\bar{K^*}$ and $\rho\rho$ are calculated as leading OZI
violation effects for $\eta_c\to \omega\phi$. We first discuss
these two processes separately, and then consider their
interferences based on the SU(3) relation.

In Table~\ref{tab-3}, the branching ratios for $\eta_c\to
\omega\phi$ are listed with the dipole form factor. It shows that
the range of $\Lambda=0.7\sim 1.2$ GeV leads to about one order of
magnitude changes to the predicted branching ratios. Although some
cautions have to be taken due to the sensitivity of the results to
the cut-off energies, we also note that the commonly adopted range
of $\Lambda=0.7\sim 1.2$ GeV has been useful for providing the
magnitude of contributions from the intermediate meson exchange
processes. We find that the intermediate $K^*\bar{K^*}$
rescattering has dominant contributions to the $\eta_c\to
\omega\phi$ transitions, while the $\rho\rho$ rescattering is more
than an order of magnitude smaller. In particular, the predicted
upper limit for the branching ratio $\eta_c\to K^*\bar{K^*}\to
\omega\phi$ is  in good agreement with the experimental upper
limit~\cite{bes-eta-c}.

As shown by the results with both $K^*\bar{K^*}$ and $\rho\rho$
contributing, the interference leads to about 2/3 enhancement to
the branching ratio with the exclusive $K^*\bar{K^*}$ rescattering
for $\Lambda=1.2$ GeV. With smaller cut-off energies, the
enhancement becomes more significant, but generally does not
change the order of magnitudes. This is an indication that for the
purpose of understanding the possible non-vanishing $\eta_c\to
\omega\phi$ decay, the estimate of the $K^*\bar{K^*}$
contributions is nearly sufficient. Certainly, more rigorous study
of the intermediate meson exchange contributions including more
meson loops should be carried out with future improved
experimental data.

In light of the improved branching ratios for $\eta_c\to
\rho\rho$, $K^*\bar{K^*}$ and $\phi\phi$, and upper limits for
$\omega\omega$ and $\omega\phi$ from BES collaboration, we
investigate the implication of OZI violations in the decays of
$\eta_c\to VV$ based on a factorization scheme recently
developed~\cite{zhao-chi-c,close-zhao-f0}. It shows that sizeable
OZI violations can occur under the present experimental accuracy.
The process of $\eta_c\to \omega\phi$ thus would be a unique test
of such effects, and we find the branching ratio is at order of
$10^{-4}$.

We then apply an intermediate meson exchange model to investigate
the OZI violation mechanism in $\eta_c\to \omega\phi$. Similar to
what was found in $J/\psi\to V f_0$ decays~\cite{zzm}, we find
that the intermediate $K^*\bar{K^*}$ rescattering can contribute
to $\eta_c\to \omega\phi$ with non-negligible magnitudes of about
 $10^{-4}$. In contrast, the contribution from the exclusive intermediate
$\rho\rho$ rescattering is relatively small. But we find that its
interference with the $K^*\bar{K^*}$ can produce sizeable effects.
Although the outputs of the intermediate meson exchange model
still bear uncertainties due to its model-dependent feature, we
find that its agreement with the factorization scheme is very
impressive. In another word, the intermediate meson exchange could
be a dynamic mechanism which leads to the OZI-rule violations in
those low-lying charmonium decays~\cite{zzm}. Nevertheless, the
factorization scheme highlights interesting features arising from
the $\eta_c$ decays, such as the correlations between the OZI rule
violations and SU(3) flavor symmetry breaking. Further experiments
at BES concentrating on $\eta_c\to \omega\phi$ should be able to
disentangle this long-standing question.

The author wishes to thank F.E. Close and B.S. Zou for useful
comments and discussions. Useful discussions with S. Jin and X.Y.
Shen are also acknowledged. This work is supported, in part, by
grants from the U.K. Engineering and Physical Sciences Research
Council (Grant No. GR/S99433/01), and the Institute of High
Energy, Chinese Academy of Sciences.


\begin{table}[ht]
\begin{tabular}{c|c|c}
\hline
Parameters & \ \ \  Fit-I \ \ \ & \ \ \ Fit-II \ \ \ \\[1ex]
\hline $r$ &  \ \ \  $0.28\pm 0.73$  \ \ \  & \ \ \   $-0.16\pm 0.15$ \ \ \   \\[1ex]\hline
$R$ & $0.83\pm 0.29$ & $1.02\pm 0.23$ \\[1ex]\hline
$g_0$ & $0.36\pm 0.04$ & $0.35\pm 0.04$\\[1ex]\hline
$\chi^2$ & $5.0\times 10^{-9}$ & $0.5$ \\[1ex]\hline
\end{tabular}
\caption{ The parameters determined in Fit-I and Fit-II. }
\label{tab-1}
\end{table}

\begin{table}[ht]
\begin{tabular}{c|c|c|c}
\hline Branching ratios $(\times 10^{-3})$ & \ \ \ BES
data~\cite{bes-eta-c} \ \ \
& \ \ \ Fit-I \ \ \ & \ \ \ Fit-II \ \ \ \\[1ex]
\hline
$\rho\rho$ & $12.5\pm 3.7\pm 5.1$ & $12.5$ & $10.7$ \\[1ex]
$K^*\bar{K^*}$ & $10.4\pm 2.6\pm 4.3$ & $10.4$ & $13.6$ \\[1ex]
$\phi\phi$ & $2.5\pm 0.5\pm 0.9$ & $2.50$ & $2.16$\\[1ex]
$\omega\omega$ &  $<6.3$ & \ \ \ $10.1$ ({\bf 10.5}) & \ \ \ $1.67$ ({\bf 1.06}) \\[1ex]
$\omega\phi$ & $<1.3 $ & \ \ \ $0.81 $ ({\bf 4.28}) & \ \ \ $0.33$ ({\bf 0.65}) \\[1ex]
\hline
\end{tabular}
\caption{ The branching ratios for $\eta_c\to VV$. The data are
from BES~\cite{bes-eta-c}. Fit-I is obtained by fitting the BES
data for $\eta_c\to \phi\phi$, $K^*\bar{K^*}$ and $\rho\rho$,
while Fit-II are obtained by including the data for $\eta_c\to
\omega\omega$ and $\omega\phi$ at the half values of their upper
limits. The bold numbers in the brackets are the root mean square
errors. } \label{tab-2}
\end{table}

\begin{table}[ht]
\begin{tabular}{c|c|c|c}
\hline
\ \ $\Lambda$ (GeV) \ \ & \ \ \  0.7 \ \ \ & \ \ \ 1.0 \ \ \ & \ \ \ 1.2 \ \ \ \\[1ex]
\hline
$K^*\bar{K^*}$ & \ \ $1.550\times 10^{-5}$ \ \  & \ \ $4.76\times 10^{-4}$ \ \ & \ \ $1.59\times 10^{-3}$ \ \  \\[1ex]\hline
$\rho\rho$ & $7.90\times 10^{-6}$ & $4.53\times 10^{-5}$ & $1.02\times 10^{-4}$ \\[1ex]\hline
$K^*\bar{K^*}+\rho\rho$ & $4.55\times 10^{-5}$ & $8.15\times
10^{-4}$ & $2.50\times 10^{-3}$ \\[1ex]\hline
\end{tabular}
\caption{ The branching ratios for $\eta_c\to \omega\phi$ with the
intermediate $K^*\bar{K^*}$, $\rho\rho$ and the sum of both at
different cut-off energies. } \label{tab-3}
\end{table}


\end{document}